\documentclass[aps,prl,twocolumn,prX,superscriptaddress]{revtex4}

\usepackage{graphicx}
\usepackage{amssymb}
\usepackage{times}
\usepackage{amsmath}
\usepackage{dcolumn}
\usepackage{bm}
\usepackage{colordvi}
\usepackage{color}

\begin{document}

\title{Observation of entanglement between two light beams spanning an octave in optical frequency}

\author{Nicolai B. Grosse} \affiliation{Quantum Optics Group,
Department of Physics, Faculty of Science, The Australian National
University, ACT 0200, Australia} 

\author{Syed Assad} \affiliation{Quantum Optics Group,
Department of Physics, Faculty of Science, The Australian National
University, ACT 0200, Australia}
\affiliation{Department of Physics, National University of Singapore, Singapore 117542}

\author{Moritz Mehmet}  \affiliation{Max-Planck-Institut f\"ur Gravitationsphysik (Albert-Einstein-Institut), Leibniz Universit\"at Hannover, Callinstra\ss e 38, 30167 Hannover, Germany}

\author{Roman Schnabel}  \affiliation{Max-Planck-Institut f\"ur Gravitationsphysik (Albert-Einstein-Institut), Leibniz Universit\"at Hannover, Callinstra\ss e 38, 30167 Hannover, Germany}

\author{Thomas Symul}  \affiliation{Quantum Optics Group,
Department of Physics, Faculty of Science, The Australian National
University, ACT 0200, Australia}

\author{Ping Koy Lam}  \affiliation{Quantum Optics Group,
Department of Physics, Faculty of Science, The Australian National
University, ACT 0200, Australia}

\begin{abstract}
We have experimentally demonstrated how two beams of light separated by an octave in frequency can become entangled after their interaction in a $\chi^{(2)}$ nonlinear medium. The entangler consisted of a nonlinear crystal placed within an optical resonator that was strongly driven by coherent light at the fundamental and second-harmonic wavelengths. An inter-conversion between the fields created quantum correlations in the amplitude and phase quadratures, which were measured by two independent homodyne detectors. Analysis of the resulting correlation matrix revealed a wavefunction inseparability of $0.74(1)<1$ thereby satisfying the criterion of entanglement. 
\end{abstract}

\date{\today}
\maketitle

\begin{figure}
\center{\includegraphics[width=\linewidth]{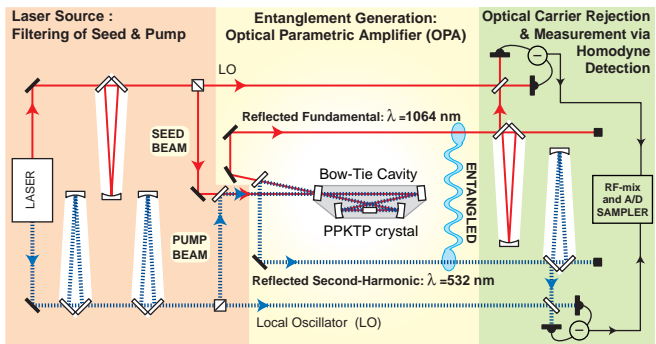}}
\caption{(color online) Entanglement generation between the reflected fundamental and second-harmonic light of an optical parametric amplifier (OPA).}
\end{figure}

Central to modern techniques in optical metrology has been the ability to make connections between light beams that span an octave in optical frequency. This development has realized the optical-comb whose offset-frequency can be directly linked to the SI definition of the second \cite{Hansch1}. As such, spectroscopic measurements can now be made with an absolute accuracy beyond one part in $10^{15}$ \cite{Hansch2}, and have enabled unprecedented testing of fundamental quantum mechanical effects \cite{Hansch3}.
Parallel to this has been progress in the area of quantum optics, where sources of non-classical light have pushed the sensitivity of spatial and temporal interferometric measurements beyond their associated classical bounds \cite{Treps&Polzik}. Recent achievements in the strength and stability of squeezed light sources suggest their imminent application to improving the sensitivity of ground-based gravitational wave detectors \cite{Schnabel1&Schnabel2&Kirk1&Caves}. Correspondingly, a source of entangled light with beams that span an octave in optical frequency and which share quantum correlations in amplitude and phase, called {\it harmonic entanglement}, has the potential to be applied to the heterodyne stabilization of optical-combs used in metrology \cite{Hansch4}. 

The goal of producing harmonic entanglement has only recently been actively pursued by experimental groups worldwide. Milestones include demonstrations of two-color discrete-variable entanglement from cascaded two-photon emission, and spontaneous parametric down-conversion \cite{Clauser&Pelton}; and two-color continuous-variable entanglement spanning $1\,{\rm nm}$ from non-degenerate optical parametric oscillators (NDOPO) \cite{Reid&Peng&Polzik&Laurat&Nussenzveig1&Peng}. The large spans offered by harmonic entanglement were predicted to occur in second-order nonlinear processes such as second-harmonic generation (SHG), above-threshold optical parametric oscillation (OPO), and optical parametric amplification (OPA) \cite{Olsen,Nussenzveig5,Lam}. 
Although a quantum correlation in amplitude between the fundamental and second-harmonic fields of SHG has been observed \cite{Zhang&Horowicz}, confirmation of entanglement also requires the detection of phase quadrature correlations of the (necessarily) bright light beams. Techniques have been used to rotate the phase quadrature onto the amplitude quadrature using either an under-coupled optical resonator, or an unequal arm length Mach-Zehnder interferometer  \cite{CavityMethod&MachZehnderMethod}. 
Yet with the most recent results from NDOPO appearing limited by excess phase noise \cite{Nussenzveig3}, a demonstration of harmonic entanglement has remained elusive.

In this Letter, we present observations of entanglement in continuous-variables between two modes of light at $532\,{\rm nm}$ and $1064\,{\rm nm}$ spanning an octave. Our best results show a wavefunction inseparability of $0.74(1)\!<\!1$ between the reflected fundamental and second-harmonic fields of a degenerate OPA. We overcame the problem of excess phase noise by choosing the driving field parameters such that the noise effectively cancelled. In addition, we demonstrated the technique of {\it optical carrier rejection} (OCR) using impedance-matched cavities, which enabled us to directly perform sideband homodyne detection on light beams that were originally an order of magnitude brighter than the local oscillator. From measurements of the amplitude and phase quadratures we determined the correlation matrix elements and tested for entanglement. We exploited the flexibility of our experimental setup to smoothly drive the system from parametric amplification to de-amplification. The results are in good agreement with a model of pump-depleted OPA.

\begin{figure}
\center{\includegraphics[width=\linewidth]{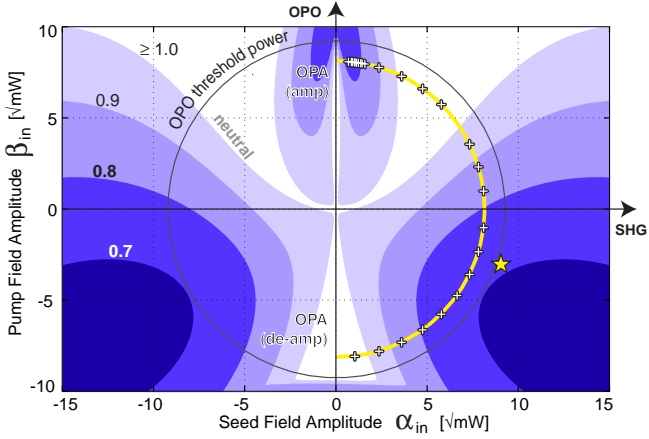}}
\caption{(color online) Map of entanglement (inseparability) as a function of seed and pump driving fields as based on our OPA model. Symbols mark where regions have been experimentally probed.}
\end{figure} 

The OPA process is described by an efficient phase-dependent exchange of energy between a fundamental {\it seed} and second-harmonic {\it pump} field. Due to this process, the light fields reflected from an OPA can be entangled in their amplitude and phase quadratures. Our aim is to obtain expressions for these observables and apply the inseparability criterion to verify entanglement. We extend our previous model \cite{Lam} to include a term that describes the effect of guided acoustic-wave Brillouin scattering (GAWBS). 
Consider a mode of light $a$ and its second-harmonic $b$, (wavelengths $\lambda_{a}\!\!=\!\!2\lambda_{b}$), which interact via a nonlinearity of strength $\epsilon$ in a single mode cavity with total decay rates $\kappa_{a,b}$. The intra-cavity fields are coupled to the environment through a mirror $\kappa_{a1,b1}$, and more weakly via other loss mechanisms $\kappa_{a2,b2}$. The system is driven by two coherent fields $\alpha_{\rm in},\beta_{\rm in}$ and can be modeled by \cite{Drummond}
\begin{eqnarray}
\dot{\hat{a}}&=& - (\kappa_{a} \!+\! i\delta w_{a})\hat{a} + \epsilon \hat{a}^{\dagger} \hat{b} + \hat{A}_{\rm in}, \\
\dot{\hat{b}}&=& - (\kappa_{b} \!+\! i\delta w_{b}) \hat{b} - \frac{1}{2}\epsilon \hat{a}^{2} + \hat{B}_{\rm in},
\label{main}
\end{eqnarray}
where input fields are denoted by $\hat{A}_{\rm in}\!=\!\sqrt{2\kappa_{a1}}\alpha_{\rm in}\! +\! \sum \sqrt{2\kappa_{a,j}}\delta\hat{A}_{j,{\rm in}}$; $\hat{B}_{\rm in}\!=\!\sqrt{2\kappa_{b1}}\beta_{\rm in}\! +\! \sum \sqrt{2\kappa_{b,j}}\delta\hat{B}_{j,{\rm in}}$ with $j\!\!\in\!\!\{1,2\}$, $\kappa_{a}\!\!=\!\kappa_{a1}\!+\kappa_{a2}$ and $\kappa_{b}\!\!=\!\kappa_{b1}\!+\kappa_{b2}$. The GAWBS noise is modeled by the detuning terms $\delta w_{a,b}\!=\!(-2\pi c/\lambda_{a,b})\xi_{a,b}\delta P$ which are driven by the dimensionless noise term $\delta P$ having variance one, but coupled via $\xi_{a,b}$; see \cite{GAWBS&Shelby&Kirk2}. We work in the Heisenberg picture where the annihilation operators $\hat{a}$ and $\hat{b}$ (and corresponding creation operators) evolve, from which the amplitude and phase quadrature operators are constructed, $\hat{X}^{+}\!\!\!=\!\!\hat{a}^{\dagger}\!+\hat{a}$ and $\hat{X}^{-}\!\!\!=\!\!i(\hat{a}^{\dagger}\!-\hat{a})$, respectively. For simplicity, we drop the hat notation henceforth. Operator linearization is used to obtain the fluctuations $(\delta X^{\pm}_{a}, \delta X^{\pm}_{b})$ centered around the classical steady-state solutions $(\alpha, \beta)$ \cite{Drummond}. 
Fourier transforming into the frequency domain allows one to solve for the driving fields in terms of the intra-cavity fields. This dependence is reversed when the equations are expressed in a matrix whose inverse is found:
\begin{equation}
\left[
\begin{array}{c}
\delta X^{+}_{a} \\
\delta X^{-}_{a} \\
\delta X^{+}_{b} \\
\delta X^{-}_{b} \\
\delta P
\end{array}
\right]
 = \left[
\begin{array}{ccccc}
A_{-} & B & C & D & F_{a}\\
B & A_{+} &-D & C & G_{a} \\
-C & D & E & 0 & F_{b} \\
-D & -C & 0 &E & G_{b} \\
0 & 0 & 0 & 0 & 1
\end{array}
\right] ^{-1}
\left[
\begin{array}{c}
\delta X^{+}_{A,{\rm in}} \\
\delta X^{-}_{A,{\rm in}} \\
\delta X^{+}_{B,{\rm in}} \\
\delta X^{-}_{B,{\rm in}}\\
\delta P^{\prime}
\end{array}
\right],
\end{equation}
where $\{ \delta X^{\pm}_{a}, \delta X^{\pm}_{b} \}$ and
$\{ \delta X^{\pm}_{A,{\rm in}}, \delta X^{\pm}_{B,{\rm in}} \}$ are the intra-cavity and accumulated input field quadratures, 
respectively;  
$A_{\pm}=\kappa_{a}-i\Omega\pm\epsilon |\beta| \cos \theta_{\beta}$, 
$B=-\epsilon |\beta | \sin \theta_{\beta}$, 
$C=-\epsilon |\alpha| \cos \theta_{\alpha}$, 
$D=-\epsilon |\alpha| \sin \theta_{\alpha}$, 
$E=\kappa_{b}-i\Omega $, 
$F_{a,b}=2i|\alpha,\beta|\sin \theta_{\alpha,\beta} (-2\pi c/\lambda_{a,b})\xi_{a,b} $, 
$G_{a,b}= 2|\alpha,\beta|\cos \theta_{\alpha,\beta} (-2\pi c/\lambda_{a,b})\xi_{a,b} $, with $\theta_{\alpha}=\textrm{Arg}(\alpha), \theta_{\beta}=\textrm{Arg}(\beta)$; and $\Omega$ the sideband frequency. The fields reflected from the resonator can be directly obtained using the input-output formalism, $\delta X^{\pm}_{{\rm
A1,ref}}\!=\! \sqrt{2 \kappa_{a1}} \delta X^{\pm}_{a}\!-\! \delta X^{\pm}_{{{\rm
A1,in}}}$; $\delta X^{\pm}_{{\rm B1,ref}}\!=\! \sqrt{2
\kappa_{b1}} \delta X^{\pm}_{b}\!-\! \delta X^{\pm}_{{{\rm B1,in}}}$  \cite{Collett}.

The resulting bi-partite Gaussian states are completely described by the correlation matrix of elements $C^{kl}_{mn}\!\!=\!\!\frac{1}{2} \langle \delta X^{k}_{m} \delta X^{l}_{n}\! +\!
\delta X^{l}_{n} \delta X^{k}_{m} \rangle$, given by
\begin{equation}
M_{ab}\!\!=\!\! \left[
\begin{array}{rrrr}
C_{aa}^{++} & C_{aa}^{+-} & C_{ab}^{++}& C_{ab}^{+-} \\
C_{aa}^{-+} & C_{aa}^{--} & C_{ab}^{-+}& C_{ab}^{--} \\
C_{ba}^{++} & C_{ba}^{+-} & C_{bb}^{++}& C_{bb}^{+-} \\
C_{ba}^{-+} & C_{ba}^{--} & C_{bb}^{-+}& C_{bb}^{--}
\end{array}
\right].
\end{equation}
where $\{ k,l\} \!\in\! \{ +,- \}$ and the the reflected field notation has been simplified with $\{ m,n \} \!\in\! \{ A1_{{\rm ref}}\!\mapsto\! a, B1_{{\rm ref}}\!\mapsto\! b \}$. We use the quantity of inseparability $\mathcal{I}$ as a measure of entanglement \cite{Duan}. It is defined from $M_{ab}$ in the following way:
\begin{eqnarray}
\mathcal{I}&=&(C_{I}^{+} + C_{I}^{-})/(2k+2/k),\label{insep1} \\
C^{\pm}_{I}&=&k C^{\pm\pm}_{aa} + (1/k)C^{\pm\pm}_{bb} - 2 |C^{\pm\pm}_{ab}|,\label{insep2}\\
k^{\pm}&=&\left[(
C^{\pm\pm}_{bb}-1)/( C^{\pm\pm}_{aa}-1)\right]^{1/2}.\label{insep3}
\end{eqnarray}
The inseparability criterion $\mathcal{I}\!\!<\!\!1$ is a necessary and sufficient condition for entanglement provided that the correlation matrix has been brought into standard form, which is obtained by applying local symplectic transformations $r_{a,b}$ to each mode separately ($\delta X_{a}^{\pm}\!\!\mapsto\!e^{\pm r_{a}}\delta X_{a}^{\pm}$ etc.) to minimize $\mathcal{I}$ and satisfy $k^{+}\!\!\!=\!\!k^{-}\!\!\!=\!\!k$; see \cite{Warwick}. Inseparability takes values in the range $[0,\infty)$ with zero signifying perfect entanglement, and unity the classical limit. 

A map of inseparability as a function of seed and pump fields is plotted in Fig.\ 2. The entangled states produced by the OPA are biased, which means that the correlations of both quadratures are not of equal strength. These biased states, however, may still be optimally entangled $(\mathcal{I}\!\rightarrow\!0)$, in the ideal limit of a lossless system \cite{Biased}. Except for the GAWBS coefficients $\xi_{a,b}$ which were fitted to the phase quadrature measurements, all parameters used in our model were obtained by characterization of the experimental setup \cite{Parameters}.

\begin{figure}
\center{\includegraphics[width=\linewidth]{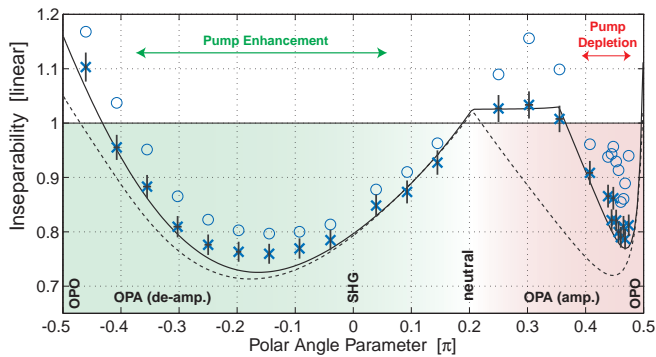}}
\caption{(color online) Inseparability as a function of angle parameter with total input power held at $65~{\rm mW}$. Entanglement is achieved for values $<\!1$. Solid and dashed curves are from our model with and without GAWBS noise, respectively. Measurements marked by (circles)crosses are (un)corrected for artifacts from the OCR process.}
\end{figure}

Figure.\ 1 shows a schematic of the experiment. Generating and measuring harmonic entanglement demands the construction of an OPA optimized for the conflicting requirements of high escape efficiency and low threshold-power. Only bright driving fields can yield strong entanglement, but the measurement of phase on bright fields is technically difficult. The source for the experiment was a Nd:YAG laser with internal frequency doubler, producing continuous-wave light at wavelengths $1064/\!\!/532\,{\rm nm}$. Where ``$/\!\!/$'' refers to parameters for each wavelength in that order, and which we call red$/\!\!/$green to aid readability. The light first needed spatial and temporal filtering via transmission through an optical cavity of linewidth $0.4\,{\rm MHz}$ for red; and two stages of filtering with $1.0\,{\rm MHz}$ then $1.9\,{\rm MHz}$ for green. Both beams were near shot-noise-limited at $7.8\,{\rm MHz}$, showing only an excess noise variance of $0.1\%$ per ${\rm mW}$ of power. A fraction of the light was diverted for the local oscillator beams (LO), before preparing the seed and pump beams with a set of RF sidebands for servo control.

The entangler itself consisted of a temperature controlled $10\,{\rm mm}$ long PPKTP crystal placed at the smaller focus of a bow-tie cavity geometry designed to minimize astigmatism. A dispersion plate placed in the larger focus allowed adjustment of the red-green phase shift. Characterization of the cavity yielded finesses of $57/\!\!/8.8$; linewidths $18/\!\!/60\,{\rm MHz}$; and reflection coefficients $74/\!\!/70\%$. The input-output mirror reflectivities were $90/\!\!/53\%$; total losses from the other mirrors were $0.6/\!\!/0.9\%$ and losses due to the crystal were $0.2/\!\!/6.7\%$ per round trip; giving an escape efficiency of $92/\!\!/86\%$. The seed$/\!\!/$pump were s-polarized and mode-matched into the cavity to $98/\!\!/99\%$. The cavity length was controlled to resonate with the seed field. The pump field was made to co-resonate by adjusting the crystal temperature. Turning the dispersion plate optimized the effective nonlinearity and lowered the OPO threshold pump power to $P_{\rm th}\!=\!85(5)\,{\rm mW}$. Note that our doubly-resonant cavity ensured a low $P_{\rm th}$ while distributing losses fairly equally over both colors. After interaction in the cavity, the reflected red and green beams (now entangled) were separated, and a $1\%$ tap-off on each was detected. This enabled locking of the seed-pump relative phase so that the OPA could operate either as an amplifier or de-amplifier.

The vital step in detection was to remove the carrier light while preserving the sidebands. This was achieved by using near impedance-matched filter cavities having linewidths $0.4/\!\!/1.0\,{\rm MHz}$. It was possible to reject $21/\!\!/26\,{\rm dB}$ of carrier light, with the cavity length locked onto resonance, and the eigenmode of the OPA spatially mode-matched to $99.9(1)/\!\!/99.9(1)\%$. Two homodyne detectors received the remaining reflected light which contained the sidebands, but also a small contribution of the original carrier light in higher order spatial modes originating from mode mismatch of the pump and seed into the OPA eigenmode.

Each red$/\!\!/$green homodyne detector had a pair of photodiodes with estimated quantum efficiency $95(2)/\!\!/95(5)\% $. The LO was mode-matched to the OPA eigenmode with visibility $99.0(2)/\!\!/99.4(2)$. The total detection efficiency was estimated to be $87/\!\!/88\%$ and  the shot-noise to dark-noise clearance was $15\,{\rm dB}$. The homodyne LO phase was servo-locked to either the phase or amplitude quadratures of the signal beam. Before taking entanglement readings, the locked quadrature angles were aligned to coincide with the max/min noise power. This method guaranteed orthogonality of the quadratures being measured, and allowed us to infer that the cross-quadrature correlations of a single color ($C_{aa}^{+-}$, etc.) were zero. We have achieved a homodyne condition with signal to LO power of $1\!\!:> \!\!30$, representing up to $26~{\rm dB}$ filtering of the carrier field whilst preserving almost all the sideband signal.

Two independent electronic channels were prepared, through which the photo-currents of the homodyne detectors were band-passed, amplified, and mixed down at $7.8\,{\rm MHz}$, before being sampled at $44.1\,{\rm kHz}$ to acquire $2.6\!\times\!10^{5}$ points. A complete data set contained paired measurements of the homodyned signal beams ($\delta X_{a}^{\pm},\delta X_{b}^{\pm}$), the signal beam without LO for OCR noise characterization, and also the vacuum noise and detector dark noise references. From this set, it was possible to normalize the quadrature amplitudes and obtain the correlation matrix elements. Corrections were made for the offsets incurred by dark noise, and excess-noise from higher-order spatial modes in the signal beam (which are artifacts of the OCR process, and not the entangler itself).

\begin{figure}
\center{\includegraphics[width=0.91\linewidth]{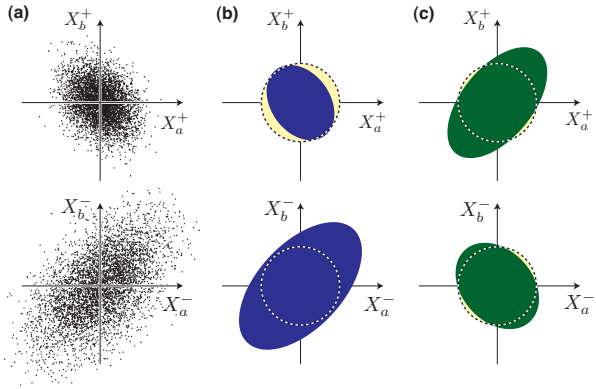}}
\caption{(color online) {\bf (a)} Time-series quadrature data showing correlations (scaled 50\%). {\bf (b)} Shaded ellipses follow a contour of the resulting probability distribution. Dashed circles mark the quantum noise limit. The quantum correlation in amplitude is evident since the ellipse falls within the circle. {\bf (c)} Dual quantum correlations are exhibited when the correlation matrix is brought into standard form.}
\end{figure}

\textit{Results:} Our aim was to drive the entangler across a range of parametric processes covering OPO, SHG, and OPA. We fixed the total driving field power to $65\,{\rm mW}$ (which is $76\%$ of $P_{\rm th}$), and adjusted the balance of power between seed and pump. In Fig.\ 3, wavefunction inseparability is plotted as a function of the polar angle of Fig.\ 2. By moving away from pure OPO, entanglement was observed over a broad range of angles $(-0.41,+0.15) \pi$, which covered parametric de-amplification through SHG, and approached the neutral point (where net inter-conversion of fundamental and second-harmonic is zero). The effect of {\it pump-enhancement} in this region was prominent. Most of the fundamental field was converted into second-harmonic field. We also found entanglement in a narrow range $(+0.41,+0.47) \pi$ corresponding to parametric amplification with weak {\it pump-depletion}. The entanglement observed in the pump-enhanced region was slightly better than in the pump-depleted region with $\mathcal{I}\!=\!0.76(2)$ and $\mathcal{I}\!=\!0.79(2)$, respectively. The dashed curve in Fig.\ 3 is a prediction based on a previous model \cite{Lam}, while the solid curve is from the extended model with GAWBS noise. The latter agrees well with the experimental results.

We observed the best harmonic entanglement in the parametric de-amplification region of strong pump-enhancement. The seed and pump powers were set at $81~{\rm mW}$ and $9~{\rm mW}$, respectively (see $\star$ in Fig.\ 2). We repeated measurements of the correlation matrix for eleven runs over many days. The ensemble average of those matrix elements in linear scale, and their $95\%$ confidence intervals based on the run-to-run variability, are presented here:
\begin{equation}
M_{ab}\!\!=\!\! \left[
\begin{array}{rrrr}
0.71(1) & 0 & -0.25(1) & -0.02(6) \\
0 & 2.45(12) & -0.07(10) & +1.42(5) \\
-0.25(1) & -0.07(10) & 0.83(2) & 0 \\
-0.02(6) & +1.42(5) & 0 & 2.56(6)
\end{array}
\right].\nonumber
\end{equation}
The matrix revealed that both colors were amplitude squeezed with $C_{aa}^{++}\!\!=\!0.71(1)$ and $C_{bb}^{++}\!\!=\!0.83(2)$. The phase quadratures showed anti-squeezing of $C_{aa}^{--}\!\!=\!2.45(12)$ and $C_{bb}^{--}\!\!=\!2.56(6)$, which imply that the Heisenberg uncertainty relation was satisfied well above the minimum uncertainty bound. These apparent ``mixed'' state statistics are a requisite of harmonic entanglement. To compute the inseparability, we performed local symplectic transformations $r_{a}=0.11(1),r_{b}=0.15(2)$ numerically to each mode such that $M_{ab}$ was brought into the standard form. Applying the inseparability definition Eq.\ (\ref{insep1}), returned a value of $\mathcal{I}=0.74(1)$, thus confirming the presence of entanglement. 

A visual representation of the correlations within the entangled state is shown in Fig.\ 4(a), where time-series quadrature data of the second-harmonic field was plotted against the fundamental field. The ellipse in (b) marks the standard deviation contour of the resulting probability distribution. The quantum anti-correlation in amplitude is evident as the ellipse falls within the circular boundary set by vacuum states. For the phase quadrature, only a classical correlation can be seen, but the proximity of the phase correlations to the classical bound is sufficient for the preservation of entanglement. This feature is symptomatic of biased entanglement \cite{Duan,Biased}. In (c) we performed local symplectic transformations to bring the correlation matrix into standard form. This led to the amplitude quadratures becoming correlated and the phase quadratures anti-correlated by an equal amount, thereby optimally redistributing the quantum correlations over both quadratures. 

In conclusion, we have observed entanglement between two light beams spanning an octave that was generated from an optical parametric amplifier. The results are in good agreement with a theoretical model that was tested across a range of parametric processes. An entanglement source of this type could aid the linking together of two optical systems at different wavelengths. In particular, bridging the octave in optical frequency between telecommunication windows and the prominent multi-level atomic systems used in quantum memories and information processing \cite{Polzik2&Kuzmich}. 

We acknowledge support from the Australian Research Council Discovery Program and the Deutsche Forschungsgemeinschaft (SFB-407).

\end{document}